\input psfig.tex
\documentclass{aa}
\usepackage{graphicx}
\begin{document}

\title{Modelling the nova rate in galaxies}
\author{Francesca Matteucci\inst{1} \and
Agostino Renda\inst{1} \and Antonio Pipino\inst{1} \and Massimo della Valle\inst{2}}
\institute{Dipartimento di Astronomia, Universit\'a di Trieste, via G.B. 
Tiepolo 11, I-34131 Trieste, Italy\\
\and Osservatorio Astronomico di Arcetri, Largo E. Fermi, Firenze, Italy}

\date{Received 2003 February 10/ Accepted 2003 April 3 }

\abstract{
We compute theoretical nova rates as well as type Ia SN rates
in galaxies of different morphological 
type (Milky Way, ellipticals and irregulars) by means of detailed chemical 
evolution models, and compare them with the 
most recent observations. The main difference among the different galaxies 
is the assumed history of star formation. 
In particular, 
we predict that the
nova rates
in giant ellipticals such as M87 are $\sim 100-300$ nova $\rm yr^{-1}$, 
about a factor of ten larger than in our 
Galaxy ($\sim 25$ nova $\rm yr^{-1}$), 
in agreement with very recent estimates from HST data. 
The best agreement
with the observed rates is obtained if the recurrence time of novae in 
ellipticals is assumed to be longer than in the Milky Way.
This result indicates that the star formation rate in ellipticals,
and in particular in M87, must have been very efficient at early 
cosmic epochs. 
We predict a nova rate for the LMC of $1.7$ nova $\rm yr^{-1}$, again 
in agreement with observations.
We compute also the K- and B-band 
luminosities for ellipticals of different luminous mass and conclude that 
there is not a clear trend for the luminosity specific nova rate with 
luminosity among these galaxies. However, firm conclusions about 
ellipticals cannot be drawn because of possible observational biases 
in observing these objects. The comparison between the specific nova 
rates in the Milky Way and the LMC 
indicates a trend of increasing nova rate passing from the Galaxy towards 
late-type spirals and Magellanic irregulars.}

\maketitle


\section{Introduction}

Classical novae are binary systems, where one component is a white
dwarf (either carbon-oxygen or neon-oxygen-magnesium), accreting
hydrogen rich material from a less evolved companion. When the pressure and
temperature at the bottom of the accreted layer exceed critical values
(e.g. Livio 1992) the nuclear reactions are ignited and novae exhibit
a sudden and rapid increase of their brightness.  The thermonuclear
runaways are responsible for the production of interesting chemical
species, such as $^{13}$C, $^{15}$N, $^{17}$O, $^{22}$Na, $^{26}$Al,
$^{22}$Ne and possibly $^{7}$Li (see Jose' \& Hernanz 1998, Gehrz et
al. 1998, Della Valle et al. 2002) and for ejection of the accreted
envelope. Although the matter processed through nova explosions is a
small fraction (a few $10^{-3}$)
of the total mass of the interstellar gas and dust in
the Milky Way, the concentrations of these rare
isotopes in the nova ejecta can be enhanced, relative to solar
abundances, by factors 10$^{2-3}$, therefore making novae non-negligible
contributors to the galactic nucleosynthesis (Romano et al. 1999,
2001; Romano \& Matteucci 2003).  Novae have been observed not only in
our Galaxy but also in the Local Group (e.g. Hubble 1929, Arp 1956 and
Rosino 1964) and beyond, in Virgo (Pritchet \& van den Bergh 1987) in M81
(Shara, Sandage \& Zurek 1999) and in Fornax (Della Valle \& Gilmozzi
2002). From these studies some estimates of nova rates have been
derived and they are shown in Table 1, where the Hubble type of the
galaxies is indicated in Col. 1, the nova rate in yr$^{-1}$ in
Col. 2, the K-band luminosity of the object in Col. 3 and in
Col. 4 the luminosity specific nova rate in units of $\rm yr^{-1}
10^{-10}L_{\odot,K}$ ($LSNR_{K}$).  These rates are taken from Della
Valle (2002) and Shafter et al. (2000). 
\begin{table*}[!ht]
\begin{flushleft}
\caption{Extragalactic nova rates and luminosity-specific nova rates from 
Della Valle (2002). The latest estimate for M87(M87*)is from 
Shara \& Zurek (2002).
With Virgo E we refer to other ellipticals in the Virgo clusters.}
\begin{tabular}{lllll}
&&&&\\
Galaxy& Hubble type& novae (yr$^{-1}$)&L$_{K}$($10^{10} L_{\odot,K}$)
& LSNR$_{K}$
(yr$^{-1}~(10^{10} L_{\odot,K})^{\rm -1}$) \\
\hline
&&&&\\
M31& Sb I-II& 29$\pm$ 4 & 18.5$\pm$ 2.4 & 1.56$\pm$ 0.30 \\
M33& Sc II-III& 4.6$\pm$ 0.9 & 1.17$\pm$0 .15& 3.93$\pm$0.93 \\
M51& Sbc I-II& 18$\pm$7& 16.6$\pm$3.4& 1.09$\pm$0.47 \\
M81& Sb I-II& 24$\pm$8& 13.3$\pm$2.7& 1.80$\pm$0.71 \\
M87& E0& 91$\pm$34& 39.8$\pm$8.2& 2.30$\pm$0.99 \\
M87$^{*}$& & 200 to 300& & \\
M101& Sc I& 12.0$\pm$4.0& 12.4$\pm$2.6& 0.97$\pm$0.38 \\
NGC5128& S0(Pec.)& 28$\pm$7& 7.94$\pm$1.10& 3.52$\pm$1.01 \\
LMC& SBm III& 2.5$\pm$0.5& 0.46$\pm$0.06& 5.42$\pm$1.30\\
SMC& Im IV-V& 0.7$\pm$0.2& 0.14$\pm$0.02& 2.19$\pm$1.49 \\
Virgo& E& 160$\pm$57& 49.7$\pm$10.5& 3.22$\pm$1.33 \\
NGC1316 & S0(Pec.)& 90 to 180& ?  & ? \\
&&&&\\
\hline
\end{tabular}
\label{Table1}
\end{flushleft}
\end{table*}

In this paper we use the chemical evolution models elaborated by
Chiappini et al. (2001), Matteucci et al. (1998) and Calura et
al. (2003) for the Milky Way, Ellipticals and late Spirals,
respectively, to compute the nova rate in `disk' dominated and `bulge'
dominated systems with the intent of checking whether
the different observed nova rates can be reproduced
under the assumption that
the main difference between galactic morphological types should 
be ascribed to the different histories of star formation.
Moreover, our study is aimed at deriving important constraints on some key
parameters of nova theory, which cannot be directly inferred from 
observations, such as the recurrence time between the outbursts, 
the fraction of binary systems which give rise
to nova systems, the delay time with which the first novae appear
relative to the beginning of star formation. The use of such models
is appropriate since
they can reproduce the basic parameters
characterizing the respective parent galaxies such as the rate of type
Ia SNe, B and K luminosities of the parent galaxies, the [$\alpha$/Fe]
ratios and global metallicities in the stellar populations and gas.
Finally, we note
that a fraction of type Ia SNe can originate from WDs in CV-like
systems, after reaching the Chandrasekhar mass limit (e.g. Livio 2000).
Whether these systems can give rise to novae or supernovae depends
crucially on the mass of the underlying white dwarf and mass accretion
rate (see Nomoto et al., 1984). Therefore, when trying to predict nova
rates in galaxies it is obvious to compute and try to reproduce the
observed SNIa rates. The {\it observed} type Ia rates in galaxies of
different morphological type are shown in Table 2. 

While the rates of
novae are expressed in $\rm yr^{-1}$ or in $LSNR_{K}$, the rates of SNIa
are expressed in $SNu_B$ (SN $\rm (100yr)^{-1}10^{-10}L_{B_{\odot}}$) and $SNu_K$,
(SN $\rm(100yr)^{-1}10^{-10}L_{K_{\odot}}$).
The paper is organized as follows: in Sect. 2 we describe how to
compute the nova and SN rates in galaxies and compare the results with
observations and finally, in Sect. 3, we draw some conclusions.

\section{Modelling the nova and supernova rates}
\subsection{The Milky Way}

The model for the Galaxy assumes two main infall episodes for the 
formation of the halo-thick disk, and the thin-disk, respectively. 
The timescale for the formation of the thin-disk is much longer 
than that of the halo, 
implying that the infalling gas forming the thin-disk comes not only from 
the halo but mainly 
from the intergalactic medium. The timescale for the formation of the thin 
disk is assumed to 
be a function of the galactocentric distance, leading to an inside-out 
picture for the Galaxy 
disk build-up, according to the original suggestion of Matteucci \& 
Fran\c cois
(1989). The two-infall model differs from other models in the 
literature mainly in two 
aspects: it considers an almost independent evolution between the halo and 
thin-disk components 
(see also Pagel \& Tautvaisiene  1995), and it assumes a threshold in
 the star 
formation process, (see Kennicutt 1989, 1998; 
van der Hulst et al. 1993;
Martin \& Kennicutt 2001). The model well reproduces the majority of 
observational constraints 
about the abundances of heavy elements both locally and in the whole disk. 

\begin{table}[!ht]
\begin{flushleft}
\caption{SNIa rate ($SNu_B$) from the combined search sample of 
Cappellaro et al. (1999) and SNIa rate in $SNu_ K$ (Della Valle \& Livio
1994) $SNu_B$ and $SNu_K$ are in units of SN
$\rm(100yr)^{-1}10^{-10}L_{B_{\odot}}$ and SN
$\rm(100yr)^{-1}10^{-10}L_{K_{\odot}}$, respectively, for
$h=0.75$.}
\begin{tabular}{lll}
&\\
Hubble type& SNeIa ($SNu_{B}$)& SNeIa ($SNu_K$)\\
\hline
&\\
E-S0& 0.18$\pm  $0.06&  0.030 $\pm 0.01$\\
Sa-Sb& 0.18$\pm $0.07&  0.045 $\pm 0.015$ \\
Sc-Sm& 0.40$\pm $0.08&  0.140 $\pm 0.03$\\
&\\
\hline
\hline
\end{tabular}
\label{Table 2}
\end{flushleft}
\end{table}

The star formation rate (SFR) adopted 
here has the same formulation as in Chiappini et al. (1997): 
\small
\begin{eqnarray}
\psi(R, t) = \nu(t)\Big(\frac{\Sigma(R,t)}{\Sigma(R_{\odot},t)} \Big)
^{2(k - 1)}\Big(\frac{\Sigma(R,t_{Gal})}{\Sigma(R,t)} \Big)^{k-1}
G_{gas}^{k}(R,t)
\end{eqnarray}
\normalsize
where $\nu(t)$ is the efficiency of the star formation process,
$\Sigma(R,t)$ is the total  
surface mass density at a given radius $R$ and given time $t$, 
$\Sigma(R_{\odot}, t)$ is the total 
surface mass density at the solar position, $G_{gas}(R, t)$ is the 
surface gas density normalized 
to the present time total surface mass density in the disk 
$\Sigma_{D}(R,t_{Gal})$, 
t$_{Gal}$ = 13 Gyr is the age of the Galaxy, R$_{\odot}$ = 8 kpc is 
the solar galactocentric 
distance 
(see Reid 1993).

The gas surface density exponent, $k$, 
is set equal to 1.5, in order to 
ensure a good fit to the observational constraints in the solar vicinity. 
This value is also 
in agreement with the observational results of Kennicutt (1998), and 
with N-body simulation 
results by Gerritsen \& Icke (1997). The star formation efficiency is set to 
$\nu$ = 2 Gyr$^{-1}$ for the Galactic halo, whereas it is $\nu$ = 1 
Gyr$^{-1}$ for the disk; 
this is to ensure the best fit to the observational features in the solar 
vicinity. The star formation
rate becomes zero when the gas surface density drops below a certain 
critical threshold. 
We adopt such a threshold density to be $(\sigma_{gas})_{th}$ $\approx$ 
4 $M_{\odot}$ 
pc$^{-2}$ for the 
Galactic halo, and $(\sigma_{gas})_{th}$ $\approx$ 7 $M_{\odot}$ pc$^{-2}$ 
for the 
disk (see Chiappini et al. 2001).  The assumption of such threshold densities
naturally produces the existence of a hiatus in the SFR
between the halo-thick disk phase and the thin-disk phase. This 
discontinuity in the SFR is observed in the [Fe/O] vs. [O/H] 
(Gratton et al. 2000) and in the [Fe/Mg] vs. [Mg/H] (Fuhrmann 1998) plots.
The IMF is that of Scalo (1986), 
assumed to be constant in 
time and space. 
We compute the nova systems formation rate ($R_{novae}$(t)) at the time $t$ 
as a 
fraction of the formation rate
 of white dwarfs at a previous time $t - \Delta t$ (see 
D'Antona \& Matteucci,  1991):
\begin{eqnarray}
R_{novae}(t)=\alpha\int_{0.8}^{8.0}\psi(t - \tau_{m} - \Delta t)\phi(m)dm.
\end{eqnarray}
Here $\Delta t$ is a delay time whose value has to be fixed to guarantee the 
cooling of the white dwarf (WD) 
to a level that ensures a strong enough nova outburst. We assume a 
distribution of 
$\Delta t$ varying from 1 to 5 Gyr. It is worth noting that assuming
a constant average delay of  $\Delta t$=2 Gyr, as assumed in previous papers
(e.g. Romano et al. 1999; 2001), produces negligible differences in the 
final results.

It is assumed that all stars with initial masses in the range 
0.8 - 8 M$_{\odot}$ end their 
lives as WDs. $\psi$(t) is the SFR as defined 
in Eq. (1), $\tau_{m}$ 
is the lifetime of the star of mass m and $\phi(m)$ is the IMF. 

The 
rate of nova eruptions is related to the WD formation rate by:
\begin{eqnarray}
R_{outbursts}(t)=\alpha R_{WDs}(t)n,
\end{eqnarray}
where $\alpha R_{WDs}=R_{novae}(t)$ is the formation rate of WDs in 
binary systems,
as defined in Eq. (2), 
which will give rise to 
nova eruptions and $n=10^{4}$ is the average number of nova outbursts for 
a typical nova system 
(Ford, 1978; Bath \& Shaviv 1978; Shara et al. 1986). This number, which is just an average value, is inversely proportional to the recurrence time.
The parameter $\alpha$, which here is a constant in space and time, can be 
fixed by reproducing
the rate of nova outbursts in our Galaxy at the present time. 
Unfortunately, estimates of this quantity in the current literature show 
a large spread. 
Predictions based on scalings from extragalactic nova surveys suggest low 
values, 
whereas estimates based on extrapolations of Galactic nova observations give 
the highest rates. 
We consider for the present time observed
rate of nova outbursts in the Galaxy 
$R_{outbursts}(t_{Gal})\approx 25\rm yr^{-1}$, as in Romano et al. (1999), for 
the following 
reasons. Given the fact that observations of novae in nearby galaxies 
would avoid, or at least 
minimize, most of 
the problems encountered by Galactic observations, such as interstellar 
extinction in the 
Galactic disk. 
Shafter (1997) shows that the nova rate based on Galactic 
observations can 
be made consistent with the rate predicted from the extragalactic data
(Della Valle et al. 1994) 
particularly if the 
Galaxy has a strong bar oriented in the direction of the Sun. In this latter, 
most favourable case, the suggested value is near $\approx 20\rm yr^{-1}$, 
otherwise, if the bar 
is weak or misaligned, the global rate can be  
$\approx 30\rm yr^{-1}$.

\begin{table}[!ht]
\begin{flushleft}
\caption{Results for the  model for the Galaxy, IMF Scalo (1986), $\alpha=0.01$, $A$=0.05, $t_{Gal}$=13 Gyr.}
\begin{tabular}{llll}
&&\\
nova (yr$^{-1}$)&  nova ($LSNR_{K}$) &  SNeIa ($100yr^{-1}$) & SNeIa (SNu)\\
\hline
&&\\
25 & 2.1 &  0.3 & 0.16\\ 
&&\\
\hline
\end{tabular}
\label{Table 3}
\end{flushleft}
\end{table}

Our first choice is to fix $\alpha$ = 0.01 (coupled with $n=10^{4}$) 
so to reproduce
the present observed nova rate of  25$\rm yr^{-1}$ and 
to use the same $\alpha$ in the models for elliptical and 
irregular galaxies. This value of $\alpha$ is the same as  in previous papers
(e.g. Romano et al. 2001; Matteucci et al. 1999).


Finally, the SNeIa rate has been computed following 
Greggio \& Renzini (1983) (see also 
Matteucci \& Greggio 1986) and is expressed as:
\begin{eqnarray}
R_{SNeIa}=A\int_{M_{B_{m}}}^{M_{B_{M}}}\phi(M_{B})\int_{\mu_{m}}^{0.5}f(\mu)\psi(t - \tau_{M_{2}})d\mu~dM_{B},
\end{eqnarray}
where $M_{2}$ is the mass of the secondary, $M_{B}$ is the total mass 
of the binary system, 
$\mu = M_{2}/M_{B}$, $\mu_{m}=max\big\{M_{2}(t)/M_{B}, 
(M_{B}-0.5M_{B_{M}})/M_{B}\big\}$, 
$M_{B_{m}}$ = 3 $\rm M_{\odot}$, $M_{B_{M}}$ = 16 $\rm M_{\odot}$. $\phi(M_{B})$ 
is the initial 
mass function for the total mass of the binary system, $f(\mu)$ is the 
distribution function 
for the mass fraction of the secondary, 
$f(\mu)=2^{1+\gamma}(1+\gamma)\mu^{\gamma}$, with 
$\gamma$ = 2;  $A$ = 0.05 is the fraction of systems with total 
mass in the appropriate range, 
which give rise to SNIa events. This quantity is fixed by reproducing the 
observed SNe Ia rate at the present epoch (see also Madau et al. 1998), 
and it provides excellent 
fits to the abundances and abundance ratios in the Galaxy.

The results for the Galaxy are shown in Table 3 and Fig. 1. 
In Table 3 we report the predicted nova and SN Ia rate for 
the Milky Way, adopting an age of 13 Gyr, according to a cosmology with 
$\Omega_m=0.3$, 
$\Omega_{\Lambda}=0.7$, $h=0.7$ and $z_f=10$. 
We computed the K- and B- band luminosity of the Milky Way by means of the 
spectro-photometric model of Jimenez et al. (1999): we find, 
for the present time, luminosities 
in very good agreement with the estimated ones.
The estimated Galactic blue luminosity is
$\rm L_{B_{Gal}}\sim 2\cdot 10^{10} L_{\odot}$ 
(see van den Bergh 1988) and the K-luminosity is 
$\rm L_{K_{GAL}} \sim 12 \cdot 10^{10}L_{\odot}$
(see Shafter el al. 2000). 
In Figure 1 we show the behaviour of the star formation rate in the 
framework of the two-infall model, where the gap between the formation of 
halo-thick disk and thin-disk is evident,  as well as the
behaviour of the nova 
and SN Ia rates. 
These latter show only a small discontinuity in correspondance with the star 
formation gap and the reason is that they both depend upon the past star 
formation.
The nova rate increases continuously up to the present time due to the 
long delay assumed for these systems (the lifetime of the WD plus the 
$\Delta t$), whereas the SNIa rate reaches a maximum and then it 
stays flat 
afterwards.

\subsection{Ellipticals}

In the models for ellipticals, the galaxies can be considered initially as an 
homogeneous 
sphere of gas with luminous masses in the range $10^{9} - 10^{12}~\rm M_{\odot}$. 
A single zone 
interstellar medium with instantaneous mixing of gas is assumed throughout.

\begin{table*}[!ht]
\caption{Model I: $x$ = 1.35, $\alpha$ = 0.01, $A$ = 0.18, t$_{gal}$ = 13Gyr.}
\begin{tabular}{llllll}
&&&&&\\
M$_{lum}$(M$_{\odot}$) & $\nu$ (Gyr$^{-1}$) & R$_{eff}$ (kpc)& t$_{GW}$ (Gyr) & L$_{K}$ ($10^{10}$L$_{K_{\odot}}$) &  L$_{B}$ ($10^{10}$L$_{B_{\odot}}$) \\
&&&&&\\
\hline
&&&&&\\
$10^9$            &  2. &   0.5   & 1.4 &  0.038 & 0.0091 \\
$10^{10}$         &  5. &   1.0   & 1.3 &  0.40 & 0.093  \\
$10^{11}$         & 11. &   3.0   & 0.6 &  5.9  & 0.70  \\ 
4$\cdot 10^{11}$  & 15. &   6.0   & 0.4 & 17.0    & 3.9   \\ 
6$\cdot 10^{11}$  & 16.5&   7.0   & 0.4 & 25.0    & 5.9    \\ 
$10^{12}$         & 20. &  10.0   & 0.4 & 45.0   & 9.6    \\
&&&&&\\
\end{tabular}
\end{table*}

\begin{table*}[!ht]
\caption{Model II:  $x$ = 0.95, $\alpha$ = 0.01, $A$ = 0.05, t$_{gal}$ = 13Gyr.}
\begin{tabular}{llllll}
&&&&&\\
M$_{lum}$(M$_{\odot}$) & $\nu$ (Gyr$^{-1}$) & R$_{eff}$ (kpc)& t$_{GW}$ (Gyr) & L$_{K}$ ($10^{10}$L$_{K_{\odot}}$) &  L$_{B}$ ($10^{10}$L$_{B_{\odot}}$) \\
&&&&&\\
\hline
&&&&&\\
$10^9$            &  2. &   0.5   & 1.4 &  0.034&  0.0044\\
$10^{10}$         &  5. &   1.0   & 0.9 &  0.42 &  0.049 \\
$10^{11}$         & 11. &   3.0   & 0.6 &  4.3  &  0.53  \\ 
4$\cdot 10^{11}$  & 15. &   6.0   & 0.4 &19.0     &  2.3   \\ 
6$\cdot 10^{11}$  & 16.5&   7.0   & 0.4 &37.0    &  3.2   \\ 
$10^{12}$         & 20. &  10.0   & 0.3 &47.7     &  6.2   \\
&&&&&\\
\end{tabular}
\end{table*}

The adopted model parameters are shown in Tables 4 and 5, 
where we show the initial 
luminous masses for the model galaxies, the assumed star formation efficiency
parameter, the effective radius, the predicted time for the occurrence 
of a galactic wind $t_{GW}$ and the predicted K- and B-luminosities. 
The parameter $A$ is varied according to the assumed IMF in order to 
reproduce the present time observed type Ia SN rate 
(Matteucci \& Gibson, 1995). On the other hand,
we kept the parameter $\alpha$ to be the same as in the Milky Way.
The model is described in
Matteucci et al. (1998) where we address the reader for more details.

The star formation
rate is given by:
\begin{eqnarray}
\psi(t)=\nu\rho_{gas}(t)/\rho(0),
\end{eqnarray}  
i.e. normalized to the initial total volume density. $\psi(t)$ is 
assumed to drop to 0 at the 
onset of the galactic wind. The quantity $\nu$ is expressed in units of
$\rm Gyr^{-1}$ and represents 
the efficiency of star formation, namely the inverse of the time scale 
of star formation. 
The star formation is assumed to stop after the development of a galactic wind
occurring before than 1 Gyr, from the beginning of star formation,
for all the galaxies listed above. 
Therefore, the star formation rate in these galaxies can be considered as a 
strong burst which does not last more than 1 Gyr and is shorter in more 
massive systems.
This is obtained by assuming that the star formation efficiency 
increases with galactic mass thus producing an `` inverse wind''
effect, as described in Matteucci (1994), 
where the galactic wind occurs before in more massive than in less massive 
ellipticals. As a consequence of this, the star formation period is longer in 
smaller systems thus allowing the SNe Ia to substantially pollute the ISM.
This effect can explain
the observed increase of the [$\alpha$/Fe] ratio with galactic mass 
(Worthey et al. 1992;
Matteucci 1994), which is not obtained in a classic wind scenario, where the 
more massive objects form stars for a longer time (Larson, 1974). 
The galactic wind develops as a consequence of the energy 
transfer from SNe into the ISM.
In fact, when the thermal energy of the gas becomes larger than the binding 
energy of the gas, the wind starts (Arimoto \& Yoshii, 1987;
Matteucci \& Tornambe' 1987; Matteucci 1992, 1994; Pipino et al. 2002).
In order to compute the binding energy of the gas some assumptions have 
to be made about the galactic potential well.
In particular,
it is assumed that all ellipticals possess heavy but diffuse dark 
matter halos;
a ratio between the half-light radius and the radius of the dark matter 
core  $R_{luminous}/R_{dark}$=0.10 and a ratio dark to luminous mass of 
10 are assumed.

The B and K luminosities for the ellipticals of different masses are 
computed by means of the spectro-photometric model of Jimenez et al. (1999) 
and are used to
compute the SN Ia rate in SNu and the nova rate per unit of $L_K$,
respectively.

The main differences between the model for the Galaxy, and the model 
for an elliptical galaxy 
concern the different SFR, which is much stronger in the earliest 
stages, 
and then is set to zero after the galactic wind in the case of ellipticals.
It is worth noting that
galactic winds in the case of elliptical galaxies
seem necessary to explain their lack of gas and the chemical enrichment 
of the ICM, whereas the Galactic model does not take into 
account the occurrence of a strong wind, mainly because of the strong 
gravitational potential 
well associated with our Galaxy, and 
the presence of the gas in the Galactic disk.

\begin{table*}[!ht]
\caption{Results for model I: nova and SN Ia rates.}
\begin{tabular}{llllll}
&&&&\\
M$_{lum}$(M$_{\odot}$)& novae (yr$^{-1}$)& novae (LSNR$_{K}$)& SNe Ia (100yr$^{-1}$)& SNe Ia ($SNu_B$)& SNe Ia ($SNu_K$)\\
&&&&\\
\hline
&&&&\\
$10^9$            &   0.6&  15.8    &     0.0026 &      0.28  & 0.07\\ 
$10^{10}$         &   8  &  20.0    &     0.027  &      0.29  & 0.08\\ 
$10^{11}$         &  97  &  16.4    &     0.19   &      0.27  & 0.03 \\ 
4$\cdot 10^{11}$  & 409  &  24.0    &     0.54   &      0.14  & 0.03\\ 
6$\cdot 10^{11}$  & 635  &  25.4    &     0.78   &      0.13  & 0.03\\ 
$10^{12}$         & 1121 &  24.9    &     1.00   &      0.10  & 0.02\\ 
&&&&\\
\end{tabular}
\end{table*}

\begin{table*}[!ht]
\caption{Results for model II: nova and SN Ia rates.}
\begin{tabular}{llllll}
&&&&\\
M$_{lum}$(M$_{\odot}$)& novae (yr$^{-1}$)& novae (LSNR$_{K}$)& SNe Ia (100yr$^{-1}$)& SNe Ia ($SNu_B$)& SNe Ia ($SNu_K$)\\
&&&&\\
\hline
&&&&\\
$10^9$            &  0.6 &   17.6    &     0.0012      & 0.27  & 0.03\\ 
$10^{10}$         &  7   &   16.7    &     0.011       & 0.22  & 0.03\\ 
$10^{11}$         &  89   &  20.7    &     0.11        & 0.21  & 0.02\\ 
4$\cdot 10^{11}$  & 374  &   19.7    &     0.35        & 0.15  & 0.02\\ 
6$\cdot 10^{11}$  & 585  &   15.8    &     0.51        & 0.16  & 0.014\\ 
$10^{12}$         & 1013   & 21.2    &     0.40        & 0.06  & 0.001\\ 
&&&&\\
\end{tabular}
\end{table*}

We adopt two different initial mass functions (IMFs): the Salpeter 
($x_{IMF}=1.35$) one and the Arimoto \& Yoshii (1987) 
($x_{IMF}=0.95$) one.
In fact, successful models of chemical evolution of ellipticals have shown 
that the Scalo IMF is not suitable for these galaxies (e.g.
Matteucci \& Gibson, 1995). 
\newline
Table 6 shows the results of the model I (Salpeter IMF), 
whereas Table 7 shows the results of the model II (x=0.95 IMF). 
These models have been computed by assuming the same $\alpha$ and 
the same $n$ as for the Milky Way. 

In particular, in the first column we report the initial galactic
luminous mass, in the second column the nova rate in units of
$\rm yr^{-1}$, in Col. 3 the luminosity specific nova rate , in Col. 4
the SN Ia rate in units of $\rm (100yr)^{-1}$ and in Col. 5 and 6 the SN Ia
rate in units of $SNu_B$ and $SNu_K$, respectively. 
The predicted nova rates are quite large for
massive ellipticals ranging between 300 and 1000 nova $\rm yr^{-1}$ which
is about a factor 3-10 larger than it is derived from observations. 
If these values were realistic, the discrepancy  
could be partially due to an observational bias affecting the ground
based nova surveys due to poor spatial resolution and bright limiting
magnitude. An indication in this direction comes from the nova rate of
M87 recently provided by Shara \& Zurek (2002) on the basis of HST
archive images, which is a factor 2--3 larger than previous
ground-based estimates (see Table 1). Another possibility is that in
early type galaxies the recurrence time between two consecutive nova
explosions is considerably longer than in late spirals as a
consequence of the different stellar population from which novae
originate. This suggestion is supported both by observational and
theoretical grounds. From one side Duerbeck (1990) and Della Valle et
al. (1992, 1994) have demonstrated, on the basis of galactic and
extragalactic nova observations, the existence of two populations of
novae: fast and bright novae belonging to `disk' stellar population
and slow and faint novae which originate from a `bulge' stellar
population. In particular, the latter authors (see also Della Valle
\& Livio 1998) suggested that bulge novae could originate from
relatively light WDs, likely in the range of masses of $\langle M_{WD}
\rangle \leq 0.9M_\odot$ while novae in the disk arise from massive
WDs ($\langle M_{WD}
\rangle \geq 1 M_\odot$). From the other side,  Truran (1990,
see also Ritter et al. 1991) has found that the mass of the WD and
the recurrence time between the outbursts are inversely proportional.
In order to match the `theoretical' with `empirical' rates, one needs
to lower either $\alpha$ or $n$ or both. For example for M87 the match
between predicted and observed rates can be achieved
by increasing the recurrence time between two consecutive outbursts, to 
$T_r\sim 3-1\cdot 10^5$ yr. This is about 10--3 times larger than
assumed for novae in the `disk' of the Galaxy.  In this way, both
the predictions of models I and II, for a luminous mass in the range
4-6 $\cdot 10^{11}M_{\odot}$ are in very good agreement with the
preliminary new estimate from Shara \& Zurek (2002) of the nova rate
for M87. In fact, M87 can be modeled both as an elliptical galaxy with
luminous mass $\sim 4\cdot 10^{11}~M_{\odot}$ and $R_{eff}\sim 6$ kpc
or with luminous mass $\sim 6\cdot 10^{11}~M_{\odot}$ and $
R_{eff}\sim$ 7 kpc (Cohen \& Ryzhov 1997).  The predictions of these
two models do not differ much and both of them can well represent M87.
Another possibility is to adopt a fraction of binary systems giving
rise to novae smaller than 0.01. Unfortunately, if from one side we can
justify this value for the Milky Way
because this $\alpha$ gives both the observed present time Galactic rate
and the correct yields of Lithium and
CNO isotopes (see Romano \& Matteucci, 2003),  on the other side, we
completely lack this information for ellipticals.

Finally, a comparison between tables 6 and 7 with Table 2 shows a very
good agreement between the observed and predicted type Ia SN rates both in units of $SNu_B$ and $SNu_K$.

In figures 2 and 3 we show the predicted star formation, nova and SNIa rates 
as functions of the cosmic time for ellipticals of different luminous 
masses and for the two IMFs.
 
Figure 4 shows the nova rate in the ellipticals as a function of the 
mass of the galaxy. From this figure one can infer the expected nova rate 
as a function of luminous galactic mass.

\subsection{Magellanic Irregulars}
As an example of Magellanic irregular galaxies  
and late-type spiral we computed a model to
fit the LMC.  In particular, we assumed an history of star formation
with two main bursts occurring during the first 3 Gyr of the galactic
life and during the last 2 Gyr, respectively, coupled with a lower
star formation rate in between the two, as shown in Figure 5. This star
formation regime for the LMC has been suggested by Calura et al. (2003)
and it can reproduce the observed
[O/Fe] versus [Fe/H].  The star
formation rate is like in Eq. (5) with an efficiency of star formation
$\nu=0.1\rm Gyr^{-1}$ in each starburst and of $\nu=0.01\rm Gyr^{-1}$ in the
interburst period.  The IMF is the Salpeter one and the parameters
$\alpha=0.01$ and $A=0.18$. The history of star formation has been
chosen to reproduce [O/Fe] vs. [Fe/H] in the LMC (see Figure 5 and
Calura et al. 2003).  We considered a luminous mass of $\sim
10^{10}M_{\odot}$ which is suitable for the LMC (e.g. Russell \& Dopita,
1992). We adopted the same values of $\alpha$ and $n$ as for the Milky
Way and found that the predicted present time nova rate is in
excellent agreement with the observed one and is $\sim 1.7$ nova
$\rm yr^{-1}$.  The predicted B and K luminosities are $L_{B}=2 \cdot
10^{9}L_{\odot}$ and $L_{K}=0.27 \cdot 10^{10}L_{\odot}$,
respectively.  As a consequence,  we predict a nova rate in units of
$LNSR_K$ of $\sim 6$, in agreement with the data of Table 1. The
predicted type Ia SN rate is $\sim 0.31$SNu, also in very good
agreement with the data of Table 2.

\begin{figure}
\resizebox{\hsize}{!}{\includegraphics{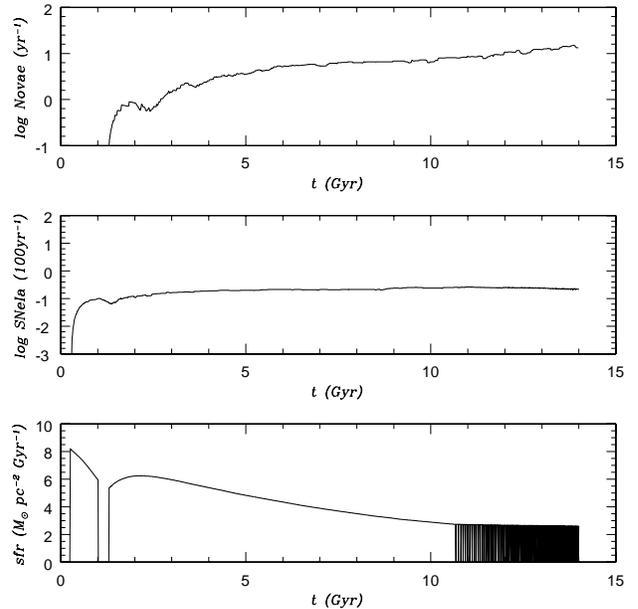}}
\caption{Nova rate and SNeIa rate predictions for the ``two-infall'' 
model for the Galaxy. Note the oscillatory behaviour of the star formation 
rate during the last 2 or 3 Gyr}
\label{Fig1}
\end{figure}

\begin{figure}
\resizebox{\hsize}{!}{\includegraphics{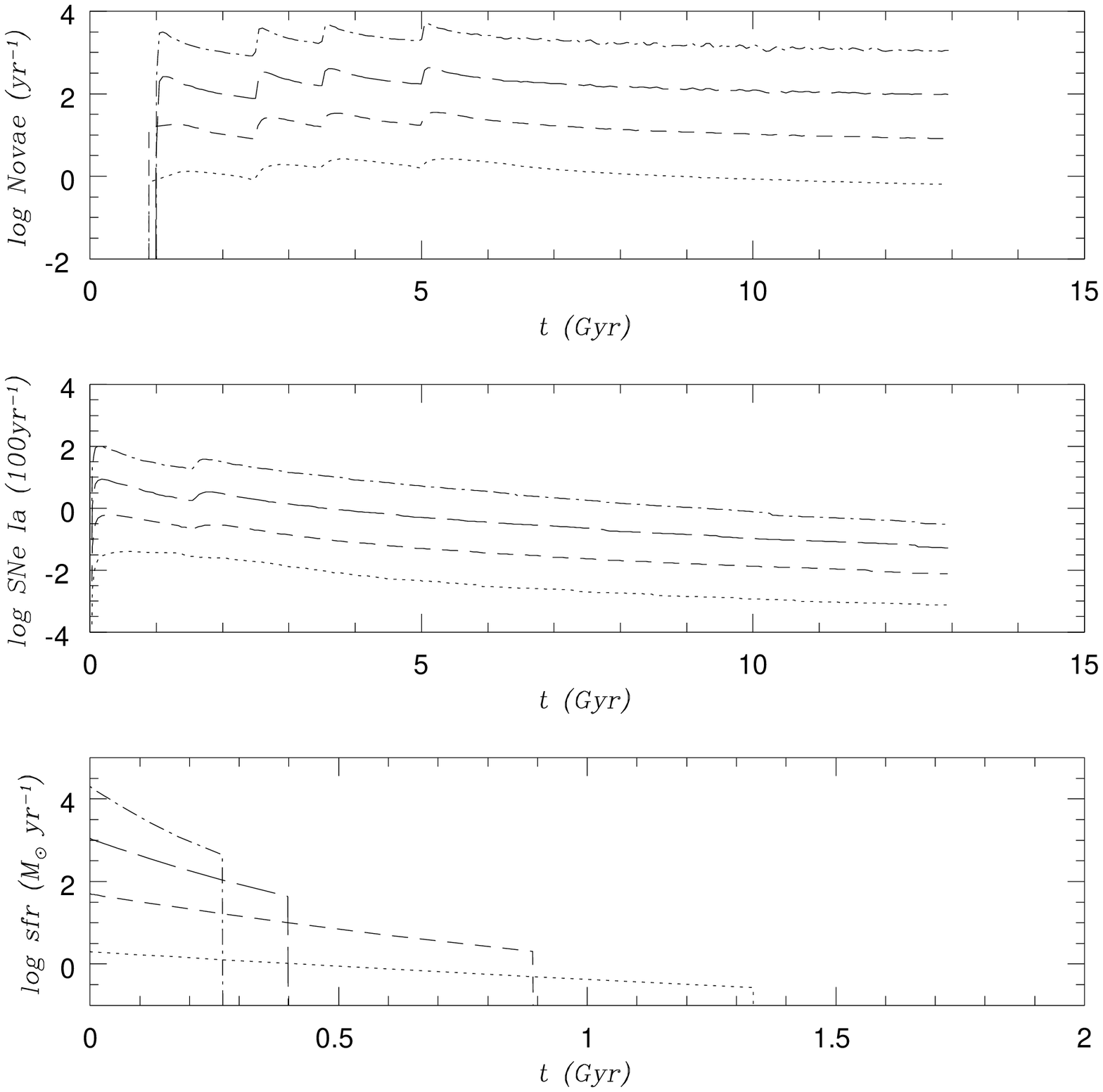}}
\caption{Ellipticals: star formation rate, SNeIa rate and novae rate predictions for 
model I. Dotted, $10^{9} M_{\odot}$; short- dashed, $10^{10} M_{\odot}$; 
long- dashed, $10^{11} M_{\odot}$; dotted - short- dashed, $10^{12} 
M_{\odot}$.}
\label{Fig2}
\end{figure}

\begin{figure}
\resizebox{\hsize}{!}{\includegraphics{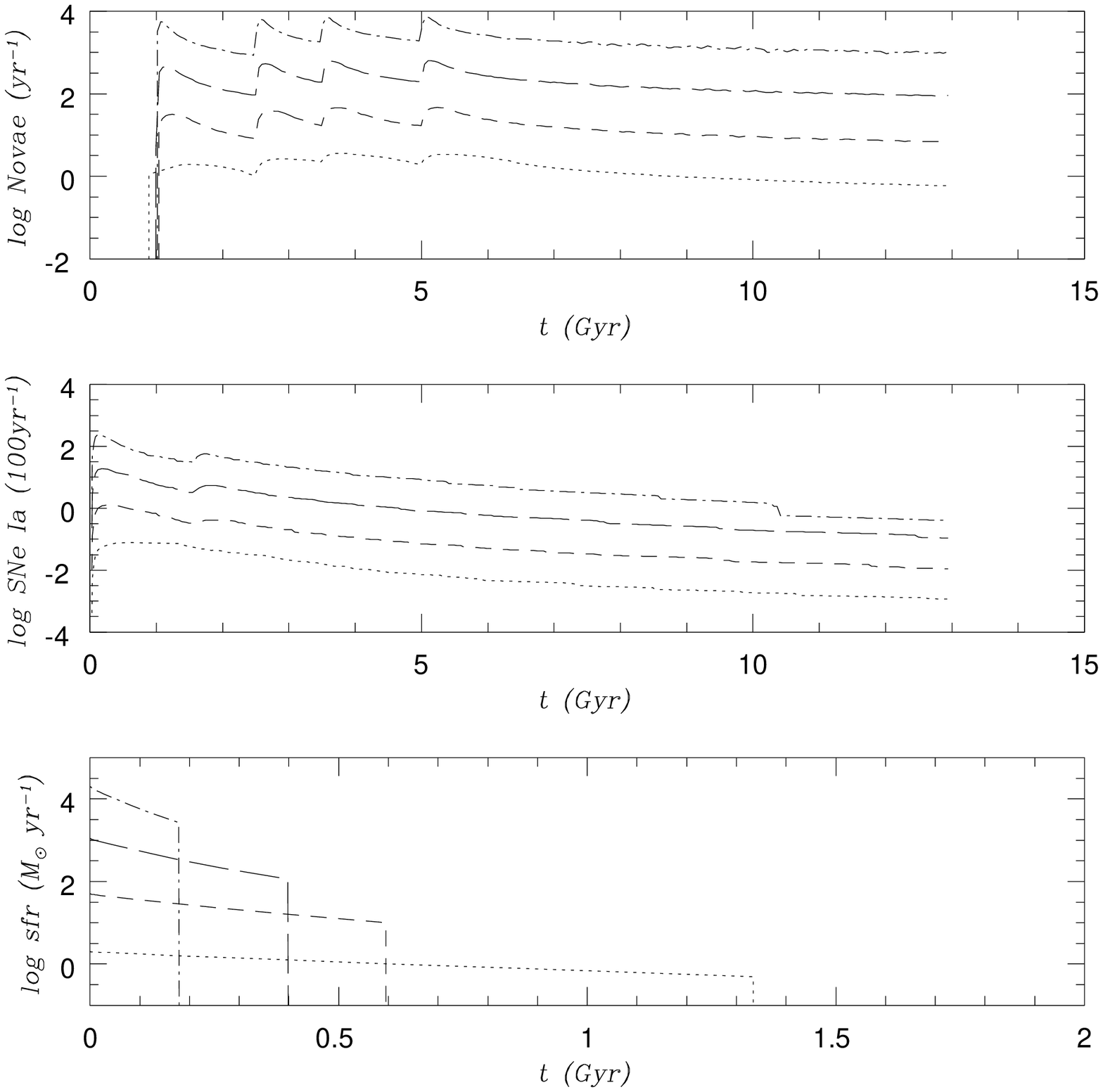}}
\caption{Ellipticals: star formation rate, SNeIa rate and novae rate predictions for model II. Dotted, $10^{9} M_{\odot}$; short- dashed, $10^{10} M_{\odot}$; long- 
dashed, $10^{11} M_{\odot}$; dotted - short- dashed, $10^{12} M_{\odot}$.}
\label{Fig3}
\end{figure}

\begin{figure}
\resizebox{\hsize}{!}{\includegraphics{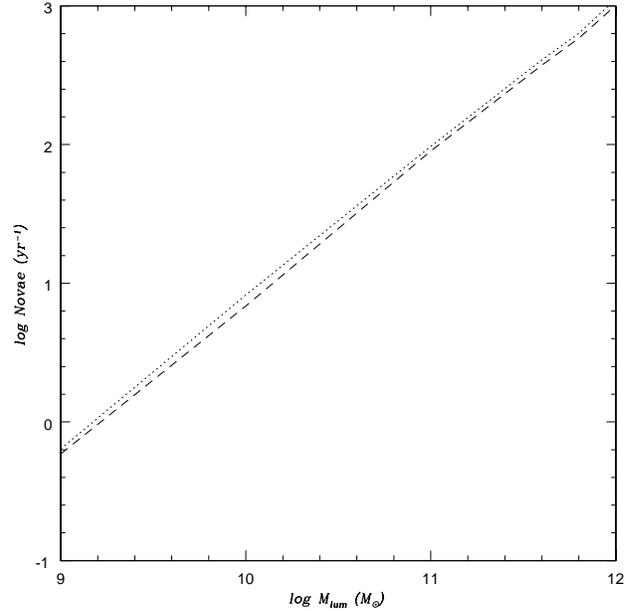}}
\caption{Ellipticals: nova rate as a function of M$_{lum}$ of the galaxy: model I 
(dotted), model II (short-dashed).} 
\label{Fig4}
\end{figure}

\begin{figure}
\resizebox{\hsize}{!}{\includegraphics{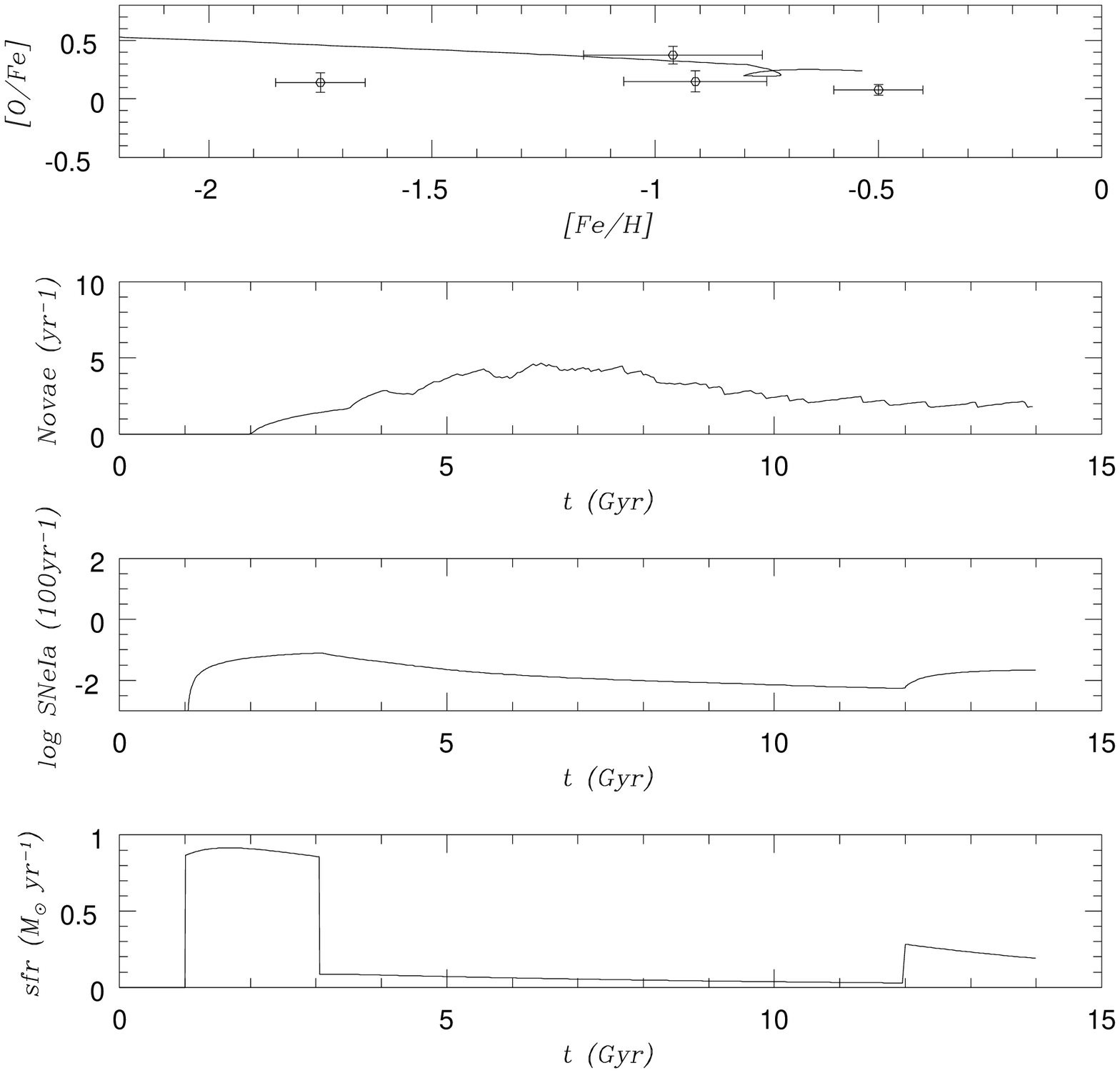}}
\caption{Irregulars: star formation rate, SNeIa rate, nova rate 
and [O/Fe] vs. [Fe/H]
predictions for the LMC. The data are from Hill et al. 2000.}
\label{Fig5}
\end{figure}

\section{Discussion and Conclusions}

In this paper we have computed the Galactic and extragalactic nova
rates, in particular the nova rates expected in elliptical galaxies
and in Magellanic irregulars, typified by the LMC. 
To do that we have
adopted detailed models of chemical and photometric evolution of the
Milky Way, ellipticals and the LMC. These models can reproduce the
majority of the observational constraints, including abundances,
abundance ratios, luminosities and colors. We have computed also the
type Ia SN rates since the novae and type Ia SNe are originating by
CV-like binary systems.  The nova rate is calculated in a simple way
but still depends on several unknown parameters such as the fraction
of binary systems which are nova progenitors (parameter $\alpha$) and
the total number of outbursts occurring during a nova lifetime
(parameter $n$).  For the sake of simplicity, we have assumed these
two parameters to be the same in all galaxies whereas the star
formation histories are different and tuned to reproduce the available
observational constraints.  For the Milky Way this particular choice
of parameters, besides reproducing the observed nova rate, allows to
predict $^{7}Li$, $^{15}N$,  $^{13}C$ and $^{17}O$, in good agreement with the
observations (Romano \& Matteucci, 2003).  On the other hand, for
external galaxies we do not have other constraints related to novae
except the nova rates.  Therefore, this approach does not allow us to
put strong constraints upon $\alpha$ and $n$ singularly but rather on
the product $\alpha \cdot n$. Concerning the delay with which novae appear 
since the beginning of star formation, we find that an average delay of 
2 Gyr 
can be assumed for all the studied galaxies.
\par
Our main conclusions can be summarized as follows:
\begin{itemize}
\item
We predict that the nova rates in ellipticals reach a maximum after 
$\sim 10^{9}$ years since the beginning of star formation and then 
they decrease
after that. The same behaviour is shared by type Ia SNe although they reach 
a maximum after only a few $10^{8}$ years 
(see also Matteucci \& Recchi, 2001). 
The delayed maximum in the nova rate relative to the SNIa rate is due to 
the longer time delay assumed for novae to appear relative to SNeIa
(see Canal et al. 1996).

\item 
Our results are suggesting that a model for ellipticals which assumes
a strong burst of star formation occurring at early cosmic times can
reproduce the majority of the observational constraints in
ellipticals.  The same model predicts that the nova rates are
increasing with the galactic mass of ellipticals and for a galaxy like
M87 we predict a present time nova rate of 300-1000 nova $\rm yr^{-1}$,
when the values of $\alpha$ and $n$ are the same as for the Galaxy.
A good agreement with the observations (100-300 nova $\rm yr^{-1}$) is
obtained if we assume an average recurrence time between two
consecutive nova outbursts of the order of $1-3 \cdot 10^5$ yr, which
is a factor of 10 larger than the typical recurrence time for novae
originating in the disk of the Milky Way. This fact is a consequence
of the existence of two populations of novae, one associated with
disk/spiral arms stellar population and the other with bulge/thick
disk stellar population (Della Valle et al., 1994; Yungelson et al. 1997). 

\item The nova rate in ellipticals is rather independent 
from the adopted IMF,
but it depends on the assumption that the efficiency of star formation 
is an increasing function of the luminous galactic mass 
(inverse wind model of Matteucci 1994). We do not predict any trend for 
the luminosity specific nova rate as a function of the galactic mass for 
these galaxies.

\item
We predict also a type Ia SN rate for ellipticals in very good agreement 
with the observed estimates and a slight decreasing trend 
for the SNIa rate with galactic blue and red luminosity.

\item
Finally, we confirm the existence of a trend for the luminosity
specific nova rate, as derived from the observations (Della Valle et
al. 1994), passing from a Magellanic irregular (spiral) like the LMC, $\sim 5$, to
a Sb spiral galaxy like the Milky Way $\sim 2.1$. For a giant
elliptical like M87 we obtain a large range of uncertainty between 2.5
and 7.5 novae/yr per unit of 10$^{10} $K luminosity. The lower limit
of the rate is very similar to that obtained from ground based nova
survey whereas the upper limit is consistent with nova rates obtained
from HST observations (Shara \& Zurek 2002). If true, this could
indicate that the studies of novae in extragalactic systems from
ground-based telescopes, can be largely affected by observational
bias. As an alternative, we note that Livio et al. (2002) have
suggested that M87 could be a fast producer of novae, because of
the presence of the jet which can enhance the accretion onto white 
dwarfs and therefore increase the nova rate.
\end{itemize}
\begin{acknowledgements}
We thank Donatella Romano for many useful suggestions and the referee
H.W. Duerbeck for his careful reading of this manuscript.
\end{acknowledgements}


\begin{thebibliography}{}
\bibitem[]{AY87} Arimoto, N., \& Yoshii, Y., 1987, A\&A, 173, 23
\bibitem[]{arp} Arp, H. C. 1956, AJ, 61, 15
\bibitem[]{BS78} Bath G.T., \& Shaviv G., 1978, MNRAS, 183, 515 
\bibitem[]{CET99} Cappellaro, E., Evans, R., \& Turatto, M., 1999, A\&A 351, 459
\bibitem[]{CMV} Calura, F., Matteucci, F., \& Vladilo, G., 2003, MNRAS in press
\bibitem[]{CRL} Canal, R., Ruiz-Lapuente, P., \& Burkert, A.,  1996 ApJ, 
456, L101
\bibitem[]{CMG97} Chiappini, C., Matteucci, F., \& Gratton, R., 1997, 
ApJ 477, 765
\bibitem[]{CMR01} Chiappini, C., Matteucci, F., \& Romano, D., 2001, 
ApJ 554, 1044
\bibitem[]{CR97}  Cohen, J. G., \& Ryzhov, A., 1997, ApJ, 486, 230    
\bibitem[]{M91} D'Antona, F., \& Matteucci, F., 1991, A\&A, 248, 62
\bibitem[]{dV02} Della Valle, M., 2002, AIP Conference Proceed. 637,
ed. M. Hernanz \& J. Isern, p. 443 
\bibitem[]{dVG02} Della Valle, M., \& Gilmozzi, M.,2002, Science, 296, 1275
\bibitem[]{DVB94} Della Valle, M., Bianchini, A., Livio, M., \& Orio, M., 
1992, A \&A, 266, 232
\bibitem[]{DLVL} Della Valle, M., \& Livio, M., 1998, ApJ, 506, 818
\bibitem []{DLV} Della Valle, M., \& Livio, M., 1994, A\&A, 286, 786
\bibitem[]{DVVW} Della Valle, M., Pasquini, L., Daou, D., \& Williams, R. E.,
2002, A\&A, 390, 155
\bibitem[]{dVetal94} Della Valle, M., Rosino, L., Bianchini, A., \& Livio, M., 
1994, A\&A 287, 403
\bibitem[]{D90} Duerbeck, H. W., 1990, I.A.U. Coll.122, eds. A. Cassatella
\& R. Viotti, Springer-Verlag: Berlin, p.34
\bibitem[]{Fo} Ford, H.C.,  1978, ApJ 219, 595
\bibitem[]{Fuh98} Furhmann, K. 1998, A\&A, 338, 161
\bibitem[]{G98} Gehrz, R. D., Truran, J. W., Williams, R. E., 
\& Starrfield, S., 1998, PASP, 110, 3
\bibitem[]{GI} Gerritsen, J. P. E., \& Icke, V., 1997, A\&A, 325, 972
\bibitem[]{Getal98} Gomez-Gomar, J., Hernanz, M., Jose, J., \& Isern, J., 1998, 
MNRAS, 296, 913
\bibitem[]{Gr01} Gratton, R., Carretta, E., Matteucci, F., \& Sneden, C., 2000,
A \& A, 358,671
\bibitem[] {GR83} Greggio, L., \& Renzini, A., 1983, A\&A, 118, 217
\bibitem[]{JPMH} Jimenez, R., Padoan, P., Matteucci, F.
\&  Heavens, A. F., 1998, MNRAS, 299, 123
\bibitem[]{JH98} Jose, J., \& Hernanz, M., 1998, ApJ, 494, 680
\bibitem[]{K89} Kennicutt, R. C., Jr., 1989, ApJ, 344, 685
\bibitem[]{K98} Kennicutt, R. C., Jr., 1998, ApJ, 498, 541
\bibitem[]{} Hill, V., Fran\c cois, P., Spite, M., Primas, F., \& Spite, F. 2000,
A \&A, 364, L19  
\bibitem[]{H} Hubble, E. P. 1929, ApJ, 69, 103
\bibitem[]{L74} Larson, R.B., 1974, MNRAS, 169, 229
\bibitem[]{L} Livio, M., 2000, in ``Type Ia SNe, Theory and Cosmology'', 
eds. J.C. Niemayer \& J.W. Truran, Cambridge Univ. Press, p.33
\bibitem[]{MDV} Madau, P. Della Valle, M., \& Panagia, N.,  1998, MNRAS, 297, L17
\bibitem[]{MK01} Martin, C. L., \& Kennicutt, R. C., Jr., 2001, ApJ, 555, 301
\bibitem[]{MF} Matteucci, F., \& Francois, P., 1989, MNRAS, 239, 885
\bibitem[]{MG86} Matteucci, F., \& Greggio, L., 1986, A\&A, 154, 279
\bibitem[]{MT87} Matteucci, F., \& Tornambe, A., 1987, A\&A, 185, 51
\bibitem[]{M92} Matteucci, F., 1992, ApJ, 397, 32
\bibitem[]{M94} Matteucci, F., 1994, A \&A, 288, 57
\bibitem[]{MG95} Matteucci, F., \& Gibson, B. K., 1995, A\&A, 304, 11
\bibitem[]{Metal98} Matteucci, F., Ponzone, R., \& Gibson, B. K., 1998, A\&A, 
335, 855
\bibitem[]{MR01} Matteucci, F., \& Recchi, S., 2001, ApJ, 558, 351
\bibitem[]{M99} Matteucci, F., Romano, D., \& Molaro, P. 1999, A \&A, 
\bibitem[]{N} Nomoto, K. Thielemann, F-K., \& Yokoi, K., 1984, ApJ, 286, 644 
\bibitem[]{PT95} Pagel, B. E. J., \& Tautvaisiene, G., 1995, MNRAS, 276, 505
\bibitem[]{PMBB} Pipino, A., Matteucci, F., Borgani, S., \& Biviano, A., 2002, 
NewA, 7, 227
\bibitem[]{PV} Pritchet, C. J., \& van den Bergh, S., 1987, ApJ, 318, 507
\bibitem[]{R93} Reid, M. J., 1993, ARAA, 31, 345
\bibitem[]{RPLW} Ritter, H., Politano, M., Livio, M., \& Webbink, R., 1991,
ApJ, 376, 177
\bibitem[]{Retal99} Romano, D., Matteucci, F., Molaro, P., \& Bonifacio, P., 
1999A\&A 352, 117
\bibitem[]{Retal01} Romano, D., Matteucci, F., Ventura, P., \& D'Antona, F., 
2001A\&A 374, 646
\bibitem[]{RM03} Romano, D., \& Matteucci, F., 2003, MNRAS in press
\bibitem[]{Ro} Rosino, L. 1964, AnAp, 27, 497
\bibitem[]{RD} Russell, S.C., \& Dopita, M.A., 1992, ApJ 384, 508
\bibitem[]{S86} Scalo, J. M., 1986, FCPh, 11, 1
\bibitem[]{Sh97} Shafter, A. W., 1997, ApJ, 487, 226
\bibitem[]{Setal00} Shafter, A. W., Ciardullo, R., \& Pritchet, C. J., 2000, 
ApJ 530, 193
\bibitem[]{Setal86} Shara, M. M., Livio, M., Moffat, A. F. J., \& Orio, M., 
1986, ApJ, 311, 163 
\bibitem[]{S00} Shara, M. M., 2000, NewAR, 44, 87
\bibitem[]{S02} Shara, M. M., \& Zurek, D.R. 2002, in ``Classical Nova 
Explosions'', ed. M. Hernanz \& J. Jos\`e, A.I.P. Conference Proceed., p.457
\bibitem[]{SSZ}Shara, M. M., Sandage, A., Zurek, D. R.,  1999, PASP, 111, 
1367
\bibitem[]{} Truran, J.W., 1990, I.A.U. Coll.122, eds. A. Cassatella
\& R. Viotti, Springer-Verlag: Berlin, p.373
\bibitem[]{vdB88} van den Bergh, S, 1988, ComAp, 12, 131
\bibitem[]{} van der Hulst, J. M., Skillman, E. D., Smith, T.,
et al. 1993, AJ, 106, 548
\bibitem[]{WFG} Worthey, G., Faber, S. M., \& Gonzalez, J. J., 1992, ApJ, 398, 69
\bibitem[]{YLT} Yungelson, L., Livio, M., \& Tutukov, A., 1997, ApJ, 481, 127
















\end{thebibliography}
\end{document}